\documentclass[a4paper,11pt]{article}
\pdfoutput=1 
\usepackage[utf8]{inputenc}
\usepackage{jheppubmod}

\usepackage{graphicx,xcolor,amsmath,amssymb,fancyhdr,siunitx,mathtools}
\usepackage[numbers]{natbib}
\usepackage[titletoc]{appendix}
\numberwithin{equation}{section}

\usepackage{cleveref}

\def\dslash{\not{\hbox{\kern-2pt $\partial$}}}
\def\Dslash{\not{\hbox{\kern-4pt $D$}}}
\def\Oslash{\not{\hbox{\kern-4pt $O$}}}
\def\Qslash{\not{\hbox{\kern-4pt $Q$}}}
\def\pslash{\not{\hbox{\kern-2.3pt $p$}}}
\def\kslash{\not{\hbox{\kern-2.3pt $k$}}}
\def\qslash{\not{\hbox{\kern-2.3pt $q$}}}
\def\svslash{\not{\hbox{\kern-2.3pt $sv$}}}
\usepackage{verbatim}

\newcommand{\md}[1]{\textcolor{blue}{[MD: #1]}}

\title{Challenges to Obtaining Results for Real QCD from SUSY QCD}

\author{Michael Dine,}

\author{Yan Yu}

\emailAdd{mdine@ucsc.edu}

\emailAdd{yanyu@ucsc.edu}

\affiliation{
    Santa Cruz Institute for Particle Physics and Department of Physics,\\University of California, Santa Cruz,\\
    Santa Cruz, CA, USA
}

\abstract{Recently there have been proposals to understand features of QCD such as confinement and chiral symmetry breaking by considering supersymmetric versions of the theory with various patterns
of soft breaking.  In this note we recall that with small soft breakings, SUSY QCD is suggestive of observed features of real QCD.  But we outline some of the challenges to establishing these features
of the theory with large soft breakings.  It appears difficult to argue for confinement or chiral symmetry breaking; at best, one can say that {\it if} the non-supersymmetric theory does not confine and/or break chiral symmetry, phase transitions would appear inevitable as one increases
the soft breakings.  We also discuss, at large N, where confinement would imply chiral symmetry breaking, the challenges to establishing confinement.}
\preprint{\begin{tabular}{l}SCIPP 18/02\\ACFI-T17-08\end{tabular}}

\begin{document}
\maketitle

\section{Introduction and Overview}

Much is understood about the dynamics of supersymmetric gauge theories.  With massless quarks, their moduli and pseudomoduli spaces have been elucidated and patterns of symmetry breaking
have been determined.  With supersymmetric masses for quarks, one can find discrete, supersymmetric ground states and determine their properties.   In certain theories which
can be obtained by deformations of $N=2$ theories, one can demonstrate confinement.   In obtaining some of these results, explicit, controlled computations are possible;
in others, holomorphy of the superpotential and gauge coupling functions are important tools.\footnote{The literature on these topics is extensive.  For reviews, with extensive references, see, for example, \cite{peskintasi,dinetextbook}.}

One might hope to break the supersymmetry and obtain results for real QCD (we will use the term ``Real QCD" for a gauge theory with three vector-like light fermions in the fundamental representation).  Adding small soft breakings is generally straightforward, and typically yields results consistent with expectations for the strong
interactions\cite{peskinsoftbreaking,hsusoftbreaking1}. \footnote{Reference \cite{komargodski} considers such issues, particularly the problem of describing
supersymmetric theories including small soft breakings in the infrared.  References \cite{shadmi,nimasoft} considered challenges to understanding breaking supersymmetry simultaneously in theories related by
Seiberg duality.} But actually breaking supersymmetry sufficiently to yield real QCD would seem challenging.  Recently it has been argued that it is possible to obtain information about QCD\cite{murayamaexactresults,murayamaconfinement} by considering a particular type of soft breaking,
 {\it anomaly mediation}\cite{anomalymediation1,anomalymediation2},  The ultraviolet insensitivity of this mechanism suggests that such an analysis might be particularly robust.

In this note we look critically at the question of establishing results in real QCD starting with the supersymmetric theory and general patterns of soft
breakings.  We recall that with small soft breakings:
\begin{enumerate}
\item  For $SU(N)$ QCD with $N_f$ flavors, one finds the symmetry breaking pattern one expects in actual QCD\cite{peskinsoftbreaking,hsusoftbreaking1}.
\item  For theories which can be obtained by deforming $N=2$ theories, either to $N=1$ supersymmetry\cite{seibergwitten1,seibergwitten2}, and then with small soft breakings to $N=0$ supersymmetry, one can demonstrate monopole condensation and confinement.
\end{enumerate}
To turn these observations into something resembling a proof of chiral symmetry breaking and/or confinement in real QCD, one would seem to need only to establish that there is a trajectory in the soft breaking parameter space
along which there is no phase transition in these theories.  One might imagine a proof by contradiction:  assume there {\it is} a phase transition to a phase with unbroken chiral symmetries or without
monopole condensation, and demonstrate some inconsistency.  But we will see that achieving such a result is a tall order,   Indeed, if one starts with the assumption
that real QCD does not break chiral symmetry and/or confine, one can advance strong arguments that there {\it is} a phase transition.  In other words, rather than being able to
rule out a phase transition to unbroken chiral symmetry or deconfinement, whether or not there is a phase transition
depends on whether real QCD does/does not break chiral symmetry/confine, i.e. on what happens in a regime where one does not have computational control over the theory.  We will see that this is the situation quite broadly, including the case of anomaly mediated supersymmetry breaking.  We will also consider
the situation at large $N$.  In large $N$, it has long been known that it is enough to establish confinement to establish the pattern of chiral symmetry breaking\cite{colemanwitten}\footnote{A recent paper elucidating aspects of the large $N$ problem is \cite{sato}.}.  We will describe
a strategy for establishing confinement at large $N$ through such deformations by considering a class of Wilsonian effective actions, but the problem of phase transitions here turns into a question
about renormalization group flows in the far infrared.

This paper is organized as follows.  In section \ref{possiblephases}, we review what it might mean to find -- or fail to find -- evidence for a phase transition (between confined and unconfined phases or chiral
symmetry broken/chiral symmetry unbroken phases) as one varies parameters.
In section \ref{susyqcdreview}, we review some aspects of
supersymmetric QCD.  We discuss theories with $N$ colors and $N_f$ flavors, the non-perturbative superpotential and patterns of supersymmetry breaking.  We discuss
confinement in $N=2$ theories broken down to $N=1$, and note that, starting with these, one can demonstrate confinement in the $N=1$ pure gauge theories with $SU(N)$ gauge
group.   In section \ref{smallsoftbreakings}, we consider the addition of general small soft breakings, with the resulting patterns of chiral symmetry breaking and the problem of confinement.  We turn to the realistic problem of large breakings in section \ref{largebreakings}.  
We consider the obstacles, for various choices of soft breaking parameters, to maintaining any theoretical control as one moves to regimes of badly broken supersymmetry.  If one {\it assumes} that the theory {\it does not} confine and/or {\it does not} break its chiral symmetries, we argue that it is quite plausible that there is a phase transition.
 In section \ref{anomalymediation}, we study anomaly mediation specifically. We explain the microscopic origin of the soft breaking potential in terms of the one loop gaugino mass,
 and explain why one loses control of the computation as the mass becomes comparable to the QCD scale.  We consider the problem of confinement in the context of the large $N$ expansion in section \ref{largen}.  We outline two challenges.  First, we note that it is again quite plausible that there is a phase transition to a state of unbroken chiral symmetry.  In this case, one would also lose confinement.   Alternatively, one can focus on confinement directly.  As we noted above, the problem can be translated into one of renormalization group flows.  It is quite plausible, and one expects, that the flow is such that confinement occurs.  Proving this is likely challenging.
We summarize in section \ref{conclusions}.

Reference \cite{luzio} considers some of these issues, focussing on anomaly mediation.  The analysis is limited to infinitesimal soft breakings, where one has the greatest control.  This paper offera speculations both pro and con on the possibility
of a phase transition.  We will discuss this work in section \ref{anomalymediation}, explaining why one expects suppression of the potential and thus a phase transition is plausible as the soft masses increase.

\section{Possible Phase Structures of Softly Broken SUSY QCD}
\label{possiblephases}

Whether one can learn about symmetry breaking and confinement by deforming supersymmetric QCD to real QCD depends on whether one can rule out a phase transition as one
increases soft breaking parameters from zero to values large compared to the QCD scale.  To understand the possibilities, we consider, first, the possible phase structure of a simple theory of a real scalar field with an exact $Z_2$ symmetry,   Suppose the potential is:
\begin{equation}
V ={1 \over 2} \mu^2 \phi^2 + {\lambda \over 4} \phi^4.
\end{equation}
Then we have:
\begin{equation}
\phi =\left  (-{\mu^2 \over \lambda} \right )^{1/2}
\end{equation}
This is singular at the origin of the parameter $\mu^2$; e.g. derivatives of $\phi$ (and other quantities) diverge as $\mu^2 \rightarrow 0$.

Add to the potential a small $Z_2$ symmetry breaking term:
\begin{equation}
V = \mu^2 \phi^2 + {\lambda \over 4} \phi^4 + \epsilon \phi.
\label{toypotential}
\end{equation}
Now there is a non-zero solution everywhere for $\phi$, and $\phi$ is a smooth function of $\mu^2$.  Derivatives of $\phi$ with respect to $\mu^2$ may be large.  For example, for $\mu^3 \gg \epsilon^{2/3} >0$ (thinking of $\lambda$ as a number of order $1$),
\begin{equation}
\phi = - {\epsilon \over \mu^2}.
\end{equation}

Here, for large soft breakings (compared to the QCD scale), $\epsilon$ would correspond to a term in the potential for the scalar fields  arising from the breaking of the chiral symmetry in real QCD.  It might
be proportional to the quark condensate:
\begin{equation}
\epsilon \propto \langle \bar q q \rangle  
\end{equation}
If $\langle \bar q q \rangle \ne 0$ in real QCD, there would not be an actual phase transition, though for large soft breakings one would find large derivatives of physical quantities in regions of the parameter space.

We might well model $\epsilon$ itself as the expectation value of another field, $\rho$, much lighter than $\phi$.  $\rho$ might have an expectation value in some regions of the parameter space,
in others not.  With small soft breakings, in particular, the degrees of freedom of supersymmetric QCD are quite different than those of ordinary QCD.  As we increase
the soft breakings, they become identical.  With large soft breakings, we can integrate out the heavy fields, potentially leading to dynamics quite different than suggested
by a simple extrapolation of the soft breaking parameters from small to large values.
As a result,
in some instances, we will be able to establish a structure of this general type, but, without solving QCD, we won't be able to determine whether $\epsilon \ne 0$.
Indeed, in the spirit of attempting a proof by contradiction, we will see that the question of symmetry breaking in QCD is precisely the question of whether $\epsilon \ne 0$; if $\epsilon = 0$, i.e. if real QCD breaks the chiral symmetry, there is no phase transition; if $\epsilon\ne 0$, there is.

We consider a model with a continuous $U(1)$ symmetry and two complex fields to illustrate some additional featues :
\begin{equation}
V = \mu^2 \vert \phi\vert^2 + {\lambda \over 4} \vert \phi \vert^4 -\epsilon \vert \rho \vert^2 + \lambda^\prime \vert  \rho \vert^4 +( \lambda^{\prime \prime} \phi \rho^3 + {\rm c.c.}  ).
\label{toypotentialwithlightscalar}
\end{equation}
with $\lambda^\prime \sim \lambda^{\prime \prime} \sim 1$.    Then in the regime where $\mu^2$ is large and positive, while $\epsilon$ is small and positive, $\rho$ has an expectation value of order $\epsilon$,
while   $\phi$ has a small expectation value, suppressed by $\epsilon \over \mu^2$.  The Goldstone boson is principally ${\rm Im}~ \rho$ (for a particular choice of phases of fields and couplings), with a small admixture (of order $\epsilon \over \mu^2$) of ${\rm Im} ~\phi$. 
The symmetry breaking dynamics are controlled by the light field, $\rho$.   Again, we might expect that as we approach real QCD from supersymmetric QCD by increasing the soft breaking masses for scalars and the gaugino, the dynamics of the quarks dominate the question of chiral symmetry breaking.
All of this is consistent with what we might expect for large soft breaking parameters in susy QCD, especially for the squark masses.  Then we can integrate out the squarks and gluinos.  The real and imaginary parts
of the squarks, in particular, will have nearly degenerate masses, and any would-be Goldstone fields must consist principally of the light fields.

Note, again, that there is no phase transition if $\epsilon$ is negative as we vary $\mu^2$ from small negative to large positive values.  If $\epsilon > 0$, there is a phase transition.  In the following sections, we will
consider how the features of this model are realized in supersymmetric QCD with various patterns of soft supersymmetry breaking, and ask similar questions about a possible deconfinement phase transition.

\section{A Brief Review of Supersymmetric Gauge Theories}
\label{susyqcdreview}

In this section we consider supersymmetric gauge theories with $N=1$ supersymmetry from the perspectives of chiral symmetry breaking; we will consider confinement
when we discuss small soft breakings in the next section.

\subsection{SUSY QCD with N colors and $N_f$ Flavors}

Here we focus on $SU(N)$ theories with $N_f$ massless flavors.  For simplicity, we consider $1 \le N_f <N$. Classically the theory has symmetry $U(N)_L \times U(N)_R \times U(1)_R$.  One linear combination of
symmetries is anomalous; the actual symmetry is
$U(1)_B \times SU(N)_L \times SU(N)_R \times U(1)_R$.
In this case, classically, there is a moduli space of zero energy states.
Up to flavor transformations,
 
\begin{equation}
Q=\bar Q =\left ( \begin{matrix}v_1 & 0 &0 & \dots \cr 0 & v_2 & 0 & \dots \cr 0 & 0 & v_3 & \dots \cr 0 & 0 & 0 & v_4 \end{matrix} \right )
\label{qvevs}
\end{equation} 
Non-perturbatively, a superspotential is generated:
\begin{equation}
W_{np} = \Lambda_{hol}^{3N - N_f \over N- N_f} {\rm det} (\bar Q Q)^{-{1 \over N-N_f}}.
\label{wnonperturbative}
\end{equation}
The resulting {\it potential} tends to zero for large $\vert v_i \vert$.
One can add quark masses, in which case there are $N$ supersymmetric vacua.

In considering the theory with soft breakings, and especially with large soft breakings, it will be helpful to have a more microscopic picture of the dynamics.  Consider, first, the case $N_f < N-1$.  The
the configuration of eqn. \ref{qvevs} breaks the gauge symmetry to $SU(N-N_f)$.  The low energy theory consists of the gauge bosons and gauginos of $SU(N-N_f)$ and a set of Goldstone bosons
and their superpartners.  These can be described by an effective lagrangian including, at low orders in $1/v$, the kinetic terms for these fields, and a coupling
\begin{equation}
{1 \over 16 \pi^2} \int d^2 \theta W_\alpha^2 \log({\rm det} (\bar Q Q)).
\label{fcoupling}
\end{equation}
corresponding to the relation of the low energy and high energy gauge couplings.  As a result, the $\Lambda$ parameter of the low energy group is related to that of the full gauge group by:
\begin{equation}
\Lambda_{LE}^3 = \Lambda^3 \left ({ \Lambda \over v} \right )^{N_f/N}. 
\end{equation}
In the low energy theory, gaugino condensation,
\begin{equation}
\langle \lambda \lambda \rangle = \Lambda_{LE}^3
\end{equation}
generates, through the coupling of equation \ref{fcoupling}, the non-perturbative superpotential.  We will have occasion to review this further in section \ref{anomalymediation}.

For the case $N_f= N-1$, the superpotential is generated by instantons\cite{ads}.  The scale invariance is broken by the scalar expectation values, and the characteristic instanton size scales with
$v$.  In slightly more detail, one integrates over a set of approximate solutions.  For very small instantons, one neglects the scalar potential\cite{thooftinstantons} and solves an equation
\begin{equation}
D^2 \phi =0,
\end{equation}
where $\phi$ stands for the $Q$ and $\bar Q$ fields and the covariant derivative refers to the gauge field configuration in the absence of the scalars, with the scale size $\rho$.  The action of the resulting configuration depends on $\rho$ as $\rho^2  v^2$, which cuts off the scale size integration, yielding contributions proportional to inverse powers of $v$ in the effective action.  The role of different scales will change as we 
add soft breakings.

This analysis represents a very safe application of ideas of effective field theory, as only the light degrees of freedom are described by this effective action.  If we turn on small soft breakings,
we we do in the next section, we will
be able to treat their effects as small perturbations of this effective lagrangian.  This will no longer be the case as the soft breakings become large.  Some degrees of freedom included in the would-be low energy theory may become heavy, while additional degrees of freedom may become light, so not only the action but even the degrees
of freedom may not be accounted for properly by the effective theory for large values of the soft breakings.

\subsection{Confinement in Theories with $N=1$ Supersymmetry Obtained by Deforming $N=2$ Theories}

It is remarkable that one can essentially prove confinement in gauge theories obtained as perturbations of $N=2$ supersymmetric theories\cite{seibergwitten1,seibergwitten2}\footnote{For a clear and concise introduction to Seiberg-Witten theory, see \cite{peskintasi}.}. The analysis starts with consideration of the moduli space of $N=2$ theories (for simplicity, without hypermultiplets).  In the language of $N=1$ supersymmetry, in addition to the gauge bosons and gauginos, there is a chiral multiplet, $\phi$, in the adjoint representation of the group.
Studying the singularities of the gauge coupling function, one can demonstrate that there are points in the moduli space of vacua at which monopole-antimonopole
pairs become massless.  For simplicity, consider the case of $SU(2)$, where
\begin{equation}
\phi = \left (\begin{matrix}v & 0 \cr 0 &-v \end{matrix} \right ).
\end{equation}
Calling the members of the monopole-antimonopole pair $M$ and $\bar M$, and
$u = {\rm Tr\phi^2}$, there is a superpotential:
\begin{equation}
W = b(u-u_0) M \bar M.
\label{monopolew}
\end{equation}
If we take $M$ and $\bar M$ to have dimension $1$, $b$ has dimension $-1$.
Now we can perturb this system, first breaking the supersymmetry to $N=1$, by introducing a mass for $\phi$, ${m_\phi \over 2} \phi^2$ in the superpotential.  (One can
see that his breaks the symmetry down to $N=1$ as the two adjoint fermions now have different
mass).  One can solve the equations
\begin{equation}
{\partial W \over \partial u} = {\partial W \over \partial M} = {\partial W \over \partial \bar M} = 0,
\end{equation}
subject to the condition that the auxiliary $D$ fields vanish for the dual group.
One has 
\begin{equation}
u = u_0;~~M = \bar M = \left ({m_\phi \over b} \right )^{1/2}.
\end{equation}
This is the phenomenon of monopole condensation.  One remarkable feature of these expressions is that we can take $m_\phi$ as large as we like, demonstrating confinement in $N=1$ supersymmetric QCD, with
no matter fields.  In making this statement, it is crucial that we know that the $N=1$ supersymmetry is unbroken.  Then we can describe the effective low energy theory
for large $\phi$ in terms of a supersymmetric effective action.  The theory possesses a a non-anomalous $R$ symmetry under which $\phi$ has charge $0$ and $m$ acts as a spurion of charge $2$. So there is only the tree level superpotential for $\phi$.  It should also be noted that the superpotential gives mass to one linear combination of $M$ and $\bar M$, while,
as a result of the gauge interactions,
the vev's for $M$ and $\bar M$ yield masses for the other linear combination and the gauge fields.  So as expected in a confining theory, the spectrum is gapped.

At first it may seem surprising that the scale of the monopole condensate grows with $m_\phi$.  But, as we elaborate further in the next subsection, the scale, $\Lambda_{QCD}$, of the low energy
gauge theory grows with $m_\phi$ as $m_{\phi}^{1/3}$.  Indeed, as we explain in the next subsection, the expectation value of the superpotential should scale with $\Lambda_{QCD}^3$
and it does.  In fact  the expectation value of the gaugino condensate matches the expectation value of $\langle W \rangle$ for the superpotential
of equation \ref{monopolew}.

We can break the supersymmetry down to $N=0$ in various ways.
If we add a small gaugino mass term, $m_\lambda$, for example, we expect that the monopole {\it potential} will be modified by terms proportional to $m_\lambda$.  For small enough $m_\lambda$, the monopoles will still condense.  Quite generally, this follows from the existence of a mass gap.  For large
$m_\lambda$, the potential is hard to compute, but we might expect it to be a complicated function of $m_\lambda$.  Absent other arguments, there is the logical possibility that at some value of $m_\lambda$, $m_\lambda = m_\lambda^0$, the condensate disappears, as does the gap in the spectrum.   We will discuss the various possibilities further shortly.


Refs. \cite{murayamaexactresults,murayamaconfinement} consider a different perturbation, which breaks the $N=2$ supersymmetric theory directly to $N=0$, motivated by ``anomaly mediation".
These authors modify only the scalar potential, adding a term:
\begin{equation}
\delta V_{am} = m \left ( {\partial W \over \partial \phi} - 3 W \right ).
\end{equation} 
This generates a minimum for the $\phi$ potential at $u = u_0$. and, again, a monopole condensate.

\subsection{Matching Gaugino Condensation and Monopole Condensation}
\label{matchingcondensates}

In the pure gauge theory, the expectation value of $W$ should be given by the gluino condensate, i.e.
\begin{equation}
\langle W \rangle = \langle \lambda \lambda \rangle = {\rm c} \Lambda_{LE}^3.
\end{equation}
Here $\Lambda_{LE}$ is the holomorphic Lambda parameter of the low energy gauge theory, a theory with, for definiteness, gauge group $SU(2)$ and no matter fields.  ${\rm c}$ is an
order one constant.

We can actually be more precise, determining the constant ${\rm c}$, following reference \cite{finnelpouliot}.  For $N_f = N-1$, these authors showed that the
non-perturbative superpotential for the massless theory, generated by instantons, is
\begin{equation}
W = {\Lambda^{2N + 1} \over {\rm det} \bar Q Q} + m \bar Q Q.
\end{equation}
Here we have added a mass term for the quarks.   Exploiting the holomorphy of the superpotential, we can take $m$ arbitrarily large.  Then we can determine that
\begin{equation}
\langle W \rangle = 2 \Lambda_{LE}^3.
\end{equation}
Here we have used:
\begin{equation}
\Lambda_{LE} = m e^{-{8 \pi^2 \over b_{LE} g^2(m)}} = m e^{-{8 \pi^2 \over g^2(M)} - {b_{HE} \over b_{LE} }\log(M/m)} = m^{1/6} \Lambda^{5/6}
\end{equation}
where $b_{LE}$ is the lowest order $\beta$ function coefficient of the low energy theory, in this case $6$, while $b_{HE}$ is the corresponding coefficient 
of the high energy $\beta$-function, here $b_{HE}=4$.  $M$ is some high energy scale, well above the expectation value of $\phi$.

Similarly, we can determine $\langle W \rangle$ from monopole condensation.  Here, following reference \cite{tachikawa}, for example, we have
\begin{equation}
u_0 = 2 \Lambda^2
\end{equation}
Then
\begin{equation}
\langle W \rangle =  m_\phi u_0 = 2 \Lambda_{LE}^2
\end{equation}
Here we have used that $\Lambda_{LE}^3 = m_\phi \Lambda^2$.
So we have complete agreement; the expectation value of the superpotential can be understood either in terms of gaugino condensation or monopole condensation.  This is quite satisfying.



Once supersymmetry is broken, the holomorphic behavior of the superpotential is lost and this connection will be modified.  Still, we have to entertain (or rule out) the possibility that if, for example, a large gaugino mass suppresses or eliminates the gluino condensate, it might also suppress or eliminate the monopole condensate.

\section{Perturbing with small soft breaking parameters}
\label{smallsoftbreakings}

To the theory with $N_f$ flavors, one can add small soft breakings, such as small masses for squarks and gauginos.  In this case, one can minimize the resulting potential for the squarks, and find stable minima at relatively large values of the fields, where one has good control over the theory (in the limit that the squark masses tend to zero, these minima must move off to infinity).  The region of small fields is selected by large mass.  This is the interesting region for studying real QCD.
We will consider the challenges to such a study shortly.

We can think of small squark masses in terms of a supersymmetric spurion field, $\Phi$, with $\Phi = \theta^2 F$, and we take $\Phi$ to have dimension $1$ (so $F$, the supersymmetry breaking
parameter, has dimension $2$).  Then the mass terms can be written as
\begin{equation}
m_{\tilde Q_{f g}}^2 \tilde Q^*_f \tilde Q_g={ \gamma_{fg} \over M^2} \int d^4 \theta \Phi^\dagger \Phi Q_f^\dagger Q_g.
\end{equation}
Note
\begin{equation}
\tilde m_Q^2 \sim {\vert F_\Phi\vert^2 \over M^2}.
\label{squarkmasses}
\end{equation}

Other supersymmetry breaking operators can be written similarly.  Consider, in particular,
\begin{equation}
{\beta \over M}  s\int  d^2 \theta \Phi W.
\end{equation}
Note that $\beta$ is a holomorphic quantity.  This gives the sort of potential correction considered by Murayama\cite{murayamaexactresults}.   At low orders in the soft breaking parameters, there are restrictions on the types of couplings due to holomorphy.  In particular, we can assign an $R$ charge to $\Phi$, $R_\Phi = -2$.  Higher powers of $\Phi$ which might appear in the effective action will be constrained by these symmetries.
The squark mass terms of equation \ref{squarkmasses}, however, are allowed.

These and similar operators may, in particular, limit the ability to constrain the theory as the soft breking masses ($F_\Phi$) become large.
But for small values of these soft breakings, we can obtain a good ground state and a definite pattern of symmetry breaking.  In particular, if we take all of the squark
masses identical, we preserve the $SU(N_f) \times SU(N_f) \times U(1)_R$ of the supersymmetric theory, and obtain a good ground state.

In more detail, in equation \ref{qvevs}, we can take $v_i = v$, a common expectation value.  The potential is now:
\begin{equation}
V = \vert {\partial W \over \partial Q_i} \vert^2 + \vert {\partial W \over \partial \bar Q_i} \vert^2 + m_{\tilde Q}^2 (\vert Q_i \vert^2 +\vert \bar Q_i \vert^2).
\end{equation}
$$~~~~\sim \Lambda^{6N-2N_f}v^{-{4N_f \over N-2N_f}-2} + m_{\tilde Q}^2 v^2.$$
So
\begin{equation}
v^2 \sim \Lambda^2 \left ( {\Lambda^2 \over m_{\tilde Q}^2} \right )^{N-N_f \over 2N}.
\end{equation}
For small $m_{\tilde Q}^2$, the system sits at large values of the fields ($v$), where the coupling is weak and the approximations are reliable.  Particularly relevant is that there is a clear separation
of mass scales.  While there are particles at the scale $v$ and particles at the scale $\Lambda$, the effective lagrangian describes particles with masses at
the much lower scale, $m_{\tilde Q}^2$.  As  $m_{\tilde Q}^2$ becomes comparable to or larger than $\Lambda$, this will no longer be the case.
The symmetry is broken to $SU(N)$.  There are a set of Goldstone bosons; their would-be superpartners have mass scaled by the supersymmetry-breaking
parameters.

Anticipating our subsequent discussion, note that if we were to continue to use this effective theory for $m_{\tilde Q}^2 \gg \Lambda^2$, the expectation values
of the squarks would be small, but the symmetries would still be broken.
This is the paradox alluded to above.  In this limit, we can integrate out the squarks.  Both the real and imaginary parts of the scalars will have large mass; any
Goldstone bosons arise from the fermionic quarks, not the squarks.

\section{What happens when the soft breakings become large?}
\label{largebreakings}

We'd like to consider combinations of soft breakings for which, as they become large, the system reduces to ordinary QCD.  We might ask:  if we assume chiral symmetry
is unbroken in real QCD, does this lead to a contradiction, once we consider the known features of supersymmetric QCD.  Similarly, if assume real QCD does not confine, does this lead to a contradiction with features of supersymmetric QCD?  The situation, as we will see, is similar to our toy model of equation \ref{toypotential}, with $\epsilon =0$.  Phase transitions
are quite plausible.

\subsection{Chiral Symmetry Breaking}

We first consider the question of chiral symmetry breaking.
Start by including masses for the squarks and the gauginos.  As these become
larger than the QCD scale, the theory indeed reduces to ordinary QCD.  These fields can be integrated out.  Indeed, examining the microscopic theory, it is the pattern of symmetry breaking at low energies
which will control the expectation values of the squarks.  How might one model the approach to this limit?

One simple approach leads to a paradox opening the possibility of a phase transition.  Suppose that one simply takes the potential arising
from the non-perturbative superpotential, equation \ref{wnonperturbative}, and adds squark masses, allowing these to be large compared to $\Lambda$.
Then for all values of the squark masses, the chiral symmetry is spontaneously broken, albeit with smaller and smaller meson decay constants.  As a result,
while we have massless Goldstone fields, we also have massive {\it real} scalar fields.  
But if the gaugino mass is large (not necessarily as large as the squark mass, but large compared to the QCD scale) this can't be.  For sufficiently large squark mass, we can integrate out the complex scalars, leaving ordinary QCD.  Whether or not there are Goldstone bosons
is a question of the dynamics of this theory.  But if there are, they are not composed of the massive scalars, but instead are the result of the (fermionic) quark dynamics.
In other words, the expectation values of the squarks may vanish, or may be small, induced by the quark condensate, as in our discussion of section \ref{possiblephases}.

To understand the issues more microscopically,
consider the class of theories with
$N > N_f + 1$.  In the supersymmetric limit, these theories have flat directions, in which the low energy theory is a pure gauge theory with gauge group $SU(N-N_f).$  Gaugino condensation in this low energy theory
generates the non-perturbative superpotential.

As we turn on soft breakings, we might expect suppression of the non-perturbative superpotential for two reasons:
\begin{enumerate}
\item  Gaugino and squark masses will suppress the coupling of the would-be modulus to the low energy gauge group.
\item  The gaugino mass will tend to suppress the size of the condensate.
\end{enumerate}

Both effects are important.  The superpotential and the non-perturbative correction to the potential, for small soft breakings, are proportional to $\langle \lambda \lambda \rangle \propto \Lambda^3$.
$\Lambda$ behaves as an inverse power of $\langle Q \rangle$ for $\langle Q \rangle > m_{\tilde Q}$.  For still lower values of the squark vev, the dependence of the condensate
on $\langle Q \rangle$ is weak.  As a reult, there is no longer enhancement of the potential for small fields.  For large soft breakings, one might expect the squark soft breaking terms
to be far larger than the non-perturbative terms in the squark potential, and the vev's to vanish at some critical value of the squark mass, modulo the
couplings of the squark fields to a {\it quark} condensate.
For large gluino mass, as we will consider shortly, one might expect further suppression of the squark potential.
In other words, as one increases the squark and gaugino masses, there is the likelihood
of a phase transition as a function of $m_\lambda$ to an unbroken phase if real QCD does not break the chiral symmetry ($\epsilon = 0$ in equation \ref{toypotential}).

 As a model for
the behavior of the condensate with $m_\lambda$, we can start with the Veneziano-Yankielowicz action\cite{vy} for the field $S = W_\alpha^2$, with a modification of the
kinetic term and with a soft breaking term for the scalar component of $S$:
\begin{equation}
{\cal L}_S = \int d^4 \theta \Lambda^{-4} S^\dagger S + \int d^2 \theta S \left ( \log (S^N/\Lambda^{3N} \right ) - N) + m_\lambda^2 \Lambda^{-4}\vert  S^2 \vert.
\end{equation}
With $m_\lambda^2 =0$, the equation ${\partial W \over \partial S} =0$ yields $S= \Lambda^3 e^{2\pi i k \over N}$, $k = 1,\dots N$, corresponding to the expected gaugino condensate, which breaks the discrete $R$ symmetry.  Adding the soft breaking term, we consider the {\it potential},
\begin{equation}
V(S, S^\dagger)  = \Lambda^4 \vert N \log (S/\Lambda^3) \vert^2 + m_\lambda^2 \Lambda^{-4} \vert S^2 \vert.
\end{equation}
The equation for the minimum of the potential is:
\begin{equation}
{\partial V \over \partial S^\dagger} = 0 = {N\Lambda^4 \log (S/\Lambda^3) \over S^\dagger}  + m_\lambda^2 \Lambda^{-4} S
\end{equation}
or
\begin{equation}
{ m_\lambda^2 \over \Lambda^2}\vert {S \over \Lambda^3} \vert^2 = -N \log(S/\Lambda^3).
\end{equation}
For small $m_\lambda$, $S$ decreases with $m_\lambda$.  This is consistent with the expectation that the non-perturbative effects responsible for the scalar potential turn off with increasing $m_\lambda$, supporting the possibility of a phase transition.

While not reliable, we can explore the behavior of this equation for large
$m_\lambda$ in order to get some sense of possible behaviors of the system.
Defining
$\hat s = S/\Lambda^3$, and $a = N m_\lambda^2/\Lambda^2$, this becomes:
\begin{equation}
\vert \hat s \vert^2 = -{1 \over a} \log \hat s.
\end{equation}
For large $a$, given that the log does not grow rapidly with $s$, we have that, roughly, 
\begin{equation}
\hat s = {{\rm number~of~order~1} \over a^{1/2}}.
\end{equation}
So for large $a$, the size of the condensate decreases as a  power of $a$.  This translates into an effective potential decreasing
as a power of $m_\lambda$.

Returning to the full theory, we now have a potential for the moduli (squark fields) behaving very differently than in the supersymmetric limit.
Instead of blowing up for small $\langle \bar Q Q \rangle$ (due to the fact that the effective coupling of the theory blows up), it now goes to zero.
Of course, we can't trust the precise form of this result -- there is no systematic treatment which justifies this analysis --but suppression of the non-perturbative effects which generated the potential for zero and small gaugino mass at large $m_\lambda$ seems almost inevitable.  Again, we note that for large squark and gaugino masses, we can integrate out these
fields, leaving only a weak dependence of the system on the expectation values of the squarks and the gaugino condensate.
These observations mean, in any case, that we can't rule out the possibility that there is a phase transition to a phase
of unbroken chiral symmetry as we now increase the soft breaking masses of the squarks, without a full understanding of symmetry breakdown in real QCD.

\subsection{$N_f = N-1$}

In the case $N_f = N-1$, the non-perturbative superpotential for zero soft mass terms is generated by instantons, which are approximate solutions of the equations of motion.  One integrates over a range of instanton scales from zero to roughly the gauge boson Compton wavelength, proportional to $\langle Q \rangle$.  The largest instantons in this range dominate.  Now as we turn on soft breakings the typical instanton
size will be less than $m_{\tilde Q}^{-1}$,   This follows by noting that for instantons of scale size $\rho < m_{\tilde Q}^{-1}$, the instanton action, $e^{-{8 \pi^2 \over g^2(\rho)}}$ no longer grows with $\rho$ as an inverse power of $\rho$, while the scale size integration involves a large positive power of $\rho$.
So just as in the case of $N_f < N-1$, we would see a significant suppression of the non-perturbative potential for large soft breakings.
Again, we would expect a phase transition unless the non-supersymmetric theory breaks the chiral symmetry by itself.  

\subsection{Breakdown of the Effective Action}

Critical to the derivation of the non-perturbative superpotential is the ability to work with an effective action written strictly in terms of the light fields.
As we make the gaugino and squark masses large compared to the QCD scale, there is no justification for keeping these fields in the effective action below
some scale slightly higher than the QCD scale. These degrees of freedom
are more massive than states of non-supersymmetric QCD which we are not including.  In section \ref{anomalymediation}, we will confront this directly.

\subsection{Confinement}

We have seen that, starting with $N=2$ theories and perturbing to $N=1$, and with small perturbations to $N=0$ (or simply perturbing to $N=0$ directly), one can demonstrate monopole
condensation and confinement.  To demonstrate confinement in theories like real QCD (here the pure gluonic theory).  Starting with the (confining) $N=1$ pure gauge theory, we need gluino masses large enough that the degrees of freedom
are just the gauge bosons.  We would expect the monopole potential to be sensitive to $m_\lambda$, in ways which are difficult to estimate.  One could well imagine that this
potential, once $m_\lambda \sim \Lambda_{QCD}$ develops a positive curvature near the origin for some critical value of the gluino mass, $m_\lambda^0$, with the
monopole condensate (and with it the mass gap associated with the dual gauge bosons) disappearing at that point.  Of course, confinement might persist and this is indeed what we expect, but it is not clear that we have the theoretical tools to assess this question.  We will consider the problem from the point of view of the
large N approximation shortly.

So it would appear, on the one hand, that it is difficult to formulate a proof of confinement perturbing the $N=2$ theory to $N=0$, while on the other hand, if one {\it assumes} that pure QCD does not confine, then a phase transition would seem inevitable.

\section{Anomaly Mediation}
\label{anomalymediation}

Reference \cite{murayamaexactresults} (and subsequently \cite{luzio}) studies a particular soft breaking pattern, referred to as ``anomaly mediation"\footnote{We put anomaly mediation in quotes
because there are really various components to this structure, including a particular, so called ``sequestered" form of the Kahler potential\cite{anomalymediation1} and
certain quantum corrections.} .
Anomaly mediation is ultraviolet insensitive.  In the present context, however, our concerns are principally with effects in the infrared, as we approach non-supersymmetric QCD in a particular manner.
We imagine specifying these parameters ini a Wilsonian effective action at scales where QCD is
weakly coupled.

\subsection{Microscopic Understanding of Anomaly Mediation in SUSY QCD with $N$ Colors and $N_f$ Flavors}

Consider the leading corrections to the potential
\begin{equation}
\delta V = m \left (\phi_i{\partial W \over \partial \phi_i} - 3 W \right )
\label{anomalymediationpotential}
\end{equation}
In the case of SUSY QCD, we should be able to understand this more microscopically.  If $N_f < N-1$, without the soft breakings, the low energy theory is a pure supersymmetric
gauge theory with a set of Goldstone supermultiplets,  The superpotential arises from gaugino condensation in this theory.  For simplicity in the writing, consider $N_f =1$.
The theory has flat directions with
\begin{equation}
Q = \bar Q = \left ( \begin{matrix} v &0 &\dots &0 \end{matrix} \right ),
\end{equation}
up to a phase.  The gauge coupling of the low energy theory depends on $v$, corresponding to an effective action:
\begin{equation}
{\cal L} = {1 \over 32 \pi^2} \left ({8 \pi^2 \over g(M)^2} + {1 \over 2} \log(\bar Q Q/M^2) \right ) W_\alpha^2.
\end{equation}
Correspondingly, in terms of the component fields, there is a term:
\begin{equation}
\langle \lambda \lambda \rangle {1 \over 64 \pi^2} \left ({F_Q \over Q} + {F_{\bar Q} \over \bar Q} \right ). 
\end{equation}
So we see that (given that $W$ is described by a single power of $\bar Q Q$:
\begin{equation}
\langle \lambda \lambda \rangle ={64\pi^2}{\partial W \over \partial Q} Q = {64 \pi^2 \over N-N_f} W.
\end{equation}

Now using the anomaly-mediated formula for the gaugino mass\cite{anomalymediation1,anomalymediation2},
\begin{equation}
m_\lambda = (3N-N_f){g^2 \over 16 \pi^2} m
\end{equation}
we can write the leading shift in the potential:
\begin{equation}
\delta V= {m_\lambda \over 2} \langle \lambda \lambda \rangle = {3 N -N_f \over 64 \pi^2 (N-N_f)} {64 \pi^2} m W.
\end{equation}
This is to be compared with the result from equation \ref{anomalymediationpotential}:
\begin{equation}
\delta V = {3 N - N_f \over N-N_f} W
\end{equation}
i.e. we have complete agreement an a microscopic understanding of equation \ref{anomalymediationpotential} in this framework.

\subsection{Implications of the Microscopic Understanding}

Note that the gaugino mass is important even though nominally a loop effect.  But note, also, that to obtain ``real QCD", we require that the gaugino mass and the squark masses
be large compared to $\Lambda_{QCD}$.\footnote{For the squark masses, the prediction for the squark masses depends on other assumptions.  For a so-called ``sequestered" Kahler
potential, the squark massess are of the same order in coupling as the gaugino masses, but they can well be larger\cite{anomalymediation1,anomalymediation2}.}  As we have seen, in this case, we expect a suppression of the gaugino condensate, and correspondingly of the superpotential.
It is then again hard to make any firm statements about the pattern of symmetry breaking, and in particular, to rule out the possibility of a phase transition.  The presence of
a phase transition or its absence
would appear to be completely contingent on the behavior of the non-supersymmetric theory.

We can ask similar questions about confinement.  One possible test is provided by large N.  For a theory with $N_f \ll N$, it is known that confinement implies chiral symmetry
breaking\cite{colemanwitten}.  The inability to establish chiral symmetry breaking than implies a similar obstacle to establishing confinement.   For theories with a single adjoint
field, we have seen that at low energies (if the adjoint is very massive)  we have a confining pure supersymmetric gauge theory.  Within anomaly mediation, the effect of supersymmetry
breaking is to give mass to the gauginos.  One cannot then readily determine what happens to the monopole condensate as the gaugino mass is increased.

\section{Large N}
\label{largen}

The large N limit potentially provides another approach to these questions.  The basic assumption of the large N limit is that the limit is smooth and that the physics of this limit is in some sense represented by a
formal sum of
the Feynman diagram expansion. One
might hope, contingent on these assumptions, to prove chiral symmetry breaking and or confinement by deforming supersymmetric systems.  In particular, confinement is a phenomenon
of the far infrared.  One might hope that a Wilsonian action, computed at some scale well below the QCD scale and the gaugino mass scale, would be characterized by features common to both the small and large gaugino mass cases.  Proving this is, in fact, the obstacle.  If one could demonstrate the existence of a mass gap in the large gaugino mass limit, one could likely prove this universality.  But this is as difficult as (and presumably equivalent to) proving confinement in any case.

In this section, we consider two aspects of the problem.   We discuss chiral symmetry breaking with small soft masses at large N.  Here we see the expected pattern of soft breakings, but, as for fixed, finite N, constructing an argument that there is no phase transition is challenging.  We consider also the problem of confinement at finite $N$ in the deformed supersymmetric CP$^N$ models.  Here there {\it is} a mass gap in both limits, and one can consider a Wilsonian action in the far infrared in each case.  One finds that these actions are universal, differing only in a numerical coefficient and leading to confinement in both cases.  We finish the section with some comments on the challenges to proving confinement in four dimensions.

\subsection{Chiral Symmetry Breaking at Large $N$}

We first consider the problem of ruling out a phase transition to unbroken chiral symmetry.
Consider, first, $N=1$ supersymmetric QCD with $N_f \ll N$ and $N \rightarrow \infty$.  With small soft breakings, elaborating slight on the work of
\cite{dinedraper}, one can show:
\begin{enumerate}
\item The $SU(N_f) \times SU(N_f)$ chiral symmetry is broken to $SU(N_f)$.
\item  $\theta$ and $\eta^\prime$ dependence (in the presence of very small quark masses) is as predicted for large $N$; physical quantities are functions of
${\theta -\eta^\prime/f_\pi \over N}$. 
\end{enumerate}
 The question is:  what happens as we turn on large soft breaking masses for squarks and gauginos. 
 Here the situation is similar to that we have encountered at finite $N$.  Ruling out a phase transition as we increase the soft breakings is a challenge.
 
 In large N we have the additional handle that, if one can prove confinement, one can prove chiral symmetry breaking in the non-supersymmetric theory\cite{colemanwitten}.
 So we turn to the question of confinement.
 

\subsection{The Problem of Confinement at Large $N$}

 We can ask the same questions about confinement.  We can consider, for example, a theory which in $N=1$ language contains a single adjoint chiral field.
 As for finite $N$, adding a supersymmetric mass term for the chiral field, one can demonstrate confinement\cite{seibergwitten1,seibergwitten2}.  This mass can be arbitrarily large, demonstrating confinement
 in the $N=1$ supersymmetric gauge theory without matter.  We will, indeed, provide further evidence for this in section \ref{matchingcondensates}.
 Then if we add small masses for the gauginos this theory confines, as it does for finite $N$.
The question becomes:  what happens as $m_\lambda$ becomes large?
Can we rule out a phase transition to a deconfined phase?

Heuristically, we might be tempted to argue that in computing, say, the expectation value of a Wilson loop for very large quark-antiquark separation, $R$, we are interested
in the behavior of the theory in the far infrared, and the Feynman diagrams are quite similar.  But if we try to be more precise, we run into obstacles.  While formally, order by
order, the diagrams have the same structure, the diagrams have a different coupling.  As the coupling is strong, it is not clear what conclusions we can draw from this formal
equivalence.  The issue is sharper if we phrase in terms of a Wilsonian effective action defined at scales below $m_\lambda$.  Framed in this way, it is not even
clear what are the appropriate degrees of freedom in the low energy theory, and what should be considered the leading operator.  Most strikingly, if real QCD did not have a mass gap,
then given that SUSY QCD does, we could have quite dramatic differences in the degrees of freedom in the low energy theory.

As we now explain, the workings of confinement in the $CP^N$ models, with and without supersymmetry, provide a possible scenario for QCD itself.

\subsection{Supersymmetric CP$^N$ As a Model }

The supersymmetric $CP^N$ theories, at large $N$, provide a model for these phenomena\cite{wittenlargeninstantons,supersymmetriccpn}.  The theories are confining
at large $N$.  It is possible to break the symmetry explicitly.  For small breaking, the confining scale is only slightly altered.  For large breaking, the purely bosonic
theory confines, as is well known.  In each case, the linear potential behaves as
\begin{equation}
V(R) = c_i \Lambda_i^2 R  ~~~i=1,2
\end{equation}
where $i=1$ is the supersymmetric case and $i=2$ is the non-supersymmetric case. $\Lambda_i$ is the lambda parameter appropriate to the two limits, and
is, in fact, identical.  In this case, there is no phase transition between the supersymmetric and non-supersymmetric phases.  The computation of the linear potential
can be described in the Wilsonian language we have used above.  The action at low energies is 
\begin{equation}
{\cal L}_{LE} = -{1 \over 4 g^2} F_{\mu \nu}^2,
\label{quadraticaction}, 
\end{equation}
with $g^2$ now a dimensionful quantity, determined by a one loop computation.   This leads to a linear potential between heavy charged objects. The quartic and higher order terms are suppressed
at small momentum transfer. 

The model is discussed in more detail in Appendix A.
But there are a few features that are instructive.  First, in the supersymmetric and non-supersymmetric theories, and points in between, the theory is gapped.  In a Wilsonian
action at scales below the
gap, the theory is described in terms of gauge degrees of freedom, $A_\mu$.  In two dimensions, these are non-dynamical, determined by the sources.  The action can be
written as a sum of terms,
\begin{equation}
{\cal L} = \sum_{n=1}^\infty {c_n \over \Lambda^{4n-2}} \left (F_{\mu \nu}^2 \right )^n
\label{cpnexpansion}
\end{equation}
In the far infrared, relevant to the heavy charged particle/antipartcle potential, all but the leading term in the sum may be ignored, and the first term gives rise to a linear potential
between these heavy objects

 We might expect that real QCD is similar.  First consider pure $N=1$ supersymmetric gauge theory.  This theory {\it we know} confines and has a mass gap.  In the far infrared, in a gauge like Coulomb gauge, we would see only non-propagating modes of the gauge field, and these would be responsible for the linear force between quarks.  These modes would be described by a simple action.   If the non-supersymmetric theory is gapped, as we expect it is, then the situation would be similar.

\section{Conclusions}
\label{conclusions}

As we have reviewed, supersymmetric QCD with small soft breakings leads to the pattern of symmetry breaking we see in nature in the strong interactions.  For theories which can be obtained as deformations of $N=2$ supersymmetric theories, one can exhibit monopole condensation and confinement.  Breaking to $N=1$, one can demonstrate this with large soft breakings, proving that, for example, a theory of gauge bosons and gauginos is confining.  Breaking to $N=0$ confinement persists with small soft breakings.  To establish that these are properties of real QCD requires proving that there is no phase transition as one gives
large masses to squarks and gauginos.  It would be enough to establish that there is no such phase transition along any trajectory  in the soft supersymmetry breaking parameter space which leads to real QCD.

We have asked whether assuming that QCD does not break chiral symmetries and/or does not confine, leads to any contradictions.  The answer to this question appears to be no; instead, we have seen that with either of these two assumptions, a phase transition would seem almost inevitable.  For the question of chiral symmetry breaking, we have argued that as one increases the gaugino mass, for example, the non-perturbative effects responsible for the interesting, theoretically controlled dynamics of susy gauge theories are suppressed.  Correspondingly, then, increasing the scalar masses, should lead to vanishing vev's for the scalar fields at some point, and correspondingly a phase transition to unbroken chiral symmetry.  For the question of confinement,
if one assumes that there is no confinement in real QCD, it is quite plausible that there is a deconfining transition as one increases soft breakings.  We have noted that this is the case
for general patterns of soft breaking, including for anomaly mediated supersymmetry breaking.

We strongly believe that real QCD breaks chiral symmetry and confines.  This because these are features of nature, which otherwise seems to be well described by QCD,
and also from results of lattice simulations and large N considerations.  As a result, as we have outlined, there is not likely to be a phase transition.
It has long been appealing to take softly broken supersymmetric gauge theories as a model for these phenomena.  
Using deformations of these theories to {\it prove} these properties of real QCD remains a daunting challenge. 

\clearpage

\appendix{\bf Appendices}

\section{\noindent The Supersymmetric CP$^N$ Model}

The CP$^N$ model, without supersymmetry, provides an interesting model for QCD\cite{wittenlargeninstantons}.  The theory is confining at large $N$, and exhibits
non-trivial $\theta$ dependence.  The model has a supersymmetric version.  This theory also is confining.  We can ask about passage from the supersymmetric to the
non-supersymmetric theory, and establish that there is no phase transition.  We'll use the notation of \cite{wittenlargeninstantons}.

\subsection{Features of the Supersymmetric and Non-Supersymmetric Models}

We'll formulate the theory as in \cite{wittenlargeninstantons}.  There are a set of bosons, $n_i$, with two dimensional Dirac fermionic partners, $\psi^i$.  At tree level
there is a gauge field, $A_\mu$, with no kinetic term, and a similar fermion, $\chi$.  The lagrangian density is:
\begin{equation}
{\cal L} = {N \over g^2} \left  [ \vert (\partial_\mu - i A_\mu) n_i \vert^2 + \bar \psi_i (i \partial_\mu - A_\mu) \gamma^\mu \psi^i+ {1 \over 2} (\sigma^2 + \pi^2)
\\  -{1 \over \sqrt{2} }\bar \psi_i (\sigma + i \pi \gamma_5) \psi^i + \lambda (\vert n_i \vert^2 -1) + \bar \chi n_i \psi^i + \bar \psi_i n^{i*} \chi. \right ]
\end{equation}

The lagrange multiplier fields $\lambda$ and $\chi$ enforce the constraints
\begin{equation}
\vert n_i \vert^2 =1;~~~\psi^i n_i = 0.
\end{equation}
The fields $A_\mu$ and $\sigma + i \pi$ are auxiliary fields, which can be eliminated from the lagrangian by solving their (algebraic) equations of motion.

In the supersymmetric case, one can compute the effective action at one loop, integrating over $\psi^i$ and $n_i$.  One can look for stationary points of $\lambda$ and
$\sigma + i \pi$.  Reference \cite{wittenlargeninstantons} finds, with ultraviolet cutoff $M$:
\begin{equation}
\lambda =   \sigma^2 ={1 \over 2} M^2 e^{-{2 \pi \over g^2}}\equiv \Lambda^2;~~~\pi= 0.
\end{equation}
As explained there, the $n_i$ and $\psi_i$ have identical, supersymmetric masses.  The gauge field and the field $\chi$ acquire non-trivial kinetic terms:
\begin{equation}
{\cal L}_{kin} = {N \over 16 \pi \Lambda^2} F_{\mu \nu}^2
\end{equation}
As a result, there is a linear force between two static charges proportional to $\Lambda^2$.

This is similar to what happens in the non-supersymmetric case, where one has only the lagrange multiplier $\lambda$ and the fields $n_i,~A_\mu$.
In fact, because the one loop determinant from the integration over the $n_i$'s is the same, the scale, $\Lambda$, which emerges is also the same.
However, the effective gauge coupling, computed at one loop, now has contributions only from bosons, not from the fermions.   This leads to a different
coefficient of the linear potential.

\subsection{Breaking Supersymmetry and Passing to the Non-Supersymmetric Theory}

In two dimensions, we can break the supersymmetry by adding mass terms for the fermions, $\psi^i$,
\begin{equation}
\delta {\cal L} = \sum_{i=1}^N m_i \psi^i \bar \psi_i.
\end{equation}
We assume that the $m_i$'s are comparable, differing by ${\cal O}(1)$ factors.
With this choice, at large $N$, the $\sigma$ vev will not significantly alter the typical masses of the fermions. For $m_i \gg \Lambda$, the fermion contributions
to the gauge boson kinetic terms become negligible; only the bosonic contribution is important.  Correspondingly, the coefficient of the linear potential, in terms of 
$\Lambda$, changes.

\section*{Acknowledgments}
We thank P. Draper and N. Seiberg for conversations.  This work was supported in part by U.S. Department of Energy grant No. DE-FG02-04ER41286.

\bibliography{real_qcd_from_susy_qcd}
\bibliographystyle{JHEP}

\end{document}